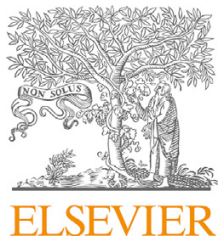
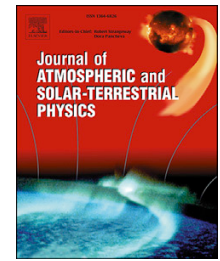

# A preliminary comparison of Na lidar and meteor radar zonal winds during geomagnetic quiet and disturbed conditions

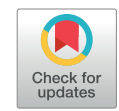

G. Kishore Kumar [a,b,*], H. Nesse Tyssøy [b], Bifford P. Williams [c]

[a] *Department of Atmospheric and Space Science, Savitribai Phule Pune University, Pune, India*
[b] *Birkeland Centre for Space Science, University of Bergen, Bergen, Norway*
[c] *GATS, Inc., Boulder, CO, USA*

ABSTRACT

We investigate the possibility that sufficiently large electric fields and/or ionization during geomagnetic disturbed conditions may invalidate the assumptions applied in the retrieval of neutral horizontal winds from meteor and/or lidar measurements. As per our knowledge, the possible errors in the wind estimation have never been reported. In the present case study, we have been using co-located meteor radar and sodium resonance lidar zonal wind measurements over Andenes (69.27°N, 16.04°E) during intense substorms in the declining phase of the January 2005 solar proton event (21–22 January 2005). In total, 14 h of measurements are available for the comparison, which covers both quiet and disturbed conditions. For comparison, the lidar zonal wind measurements are averaged over the same time and altitude as the meteor radar wind measurements. High cross correlations (~0.8) are found in all height regions. The discrepancies can be explained in light of differences in the observational volumes of the two instruments. Further, we extended the comparison to address the electric field and/or ionization impact on the neutral wind estimation. For the periods of low ionization, the neutral winds estimated with both instruments are quite consistent with each other. During periods of elevated ionization, comparatively large differences are noticed at the highermost altitude, which might be due to the electric field and/or ionization impact on the wind estimation. At present, one event is not sufficient to make any firm conclusion. Further study with more co-located measurements are needed to test the statistical significance of the result.

## 1. Introduction

Investigations of energetic particle precipitation (EPP) impact on the middle atmosphere have a long history, which starts in the late 1960s. Such studies have gained particularly strong attention in the last few decades. Energetic particles (protons, electrons, heavier ions) precipitate from different sources: directly from the Sun in large solar particle events (SPEs), from the plasma sheet and the radiation belts during geomagnetic storms and substorms, or from outside the solar system. The particles from different sources have different energy spectra and hence affect different altitudes and geographic locations (Sinnhuber et al., 2012). The EPP events can last up to a few days and lead to polar atmospheric changes through ionization, dissociation, dissociative ionization, and excitation processes. They are known to cause significant changes in chemical constituents such as $HO_x$ (H, OH, $HO_2$), $NO_x$ (N, NO, $NO_2$), and ozone, which in turn may cause changes in the associated heating and cooling rates. Changes in the temperature will impact the middle atmosphere residual circulation. The chemical changes during EPP events are evident even in small geomagnetic storms (Zawedde et al., 2016), while the subsequent potential dynamical changes are poorly understood. Detailed information on middle atmospheric chemical changes during EPP can be found in Sinnhuber et al. (2012). Very few observations have, however, reported the dynamical changes associated with EPP in the mesosphere lower thermosphere (MLT) (e.g., Pancheva et al., 2007; Singer et al., 2013; Trifonov et al., 2016).

The MLT is characterized as an ocean of dynamical changes ranging from short time scales, such as gravity waves, to large time scales, such as quasi-biennial oscillation, and their impact varies from regional response to global circulation changes. Both ground-based and space-based instruments are used to understand the MLT region. Although satellites provide global coverage, the coverage over polar latitudes is less complete. Ground-based observations such as MF radar (e.g., Manson and Meek, 1991; Kishore Kumar et al., 2014a), Meteor radar (e.g., Hocking, 2001; Pancheva and Mitchell, 2004; Pancheva et al., 2007; Singer et al., 2013; Kishore Kumar et al., 2014a, b) and lidar (She and Yu, 1994) are powerful tools and have been widely operated over different latitudes and longitudes. They provide a wealth of information about MLT dynamics. Each of these instruments has its own spatial and temporal coverage accompanied by advantages and disadvantages. The MF radar technique makes use of the ionized component of the atmosphere as a tracer for the neutral motions in the altitude region 50–110 km and provide neutral winds with a good time resolution. However, it has a limitation during strong ionization events such as EPP, which saturate the MF radar system and make it inefficient in resolving the neutral motions. The meteor radar technique, when implemented properly, can provide both wind and temperature information. It is based on the ionized column (meteor trail) created by meteor ablations. These ionized columns can strongly backscatter radar pulses in a direction at right angles to the long axis of the ionized column. By measuring the Doppler shift resulting from the motion of the meteor trail, a pulsed Doppler radar can be used to profile the neutral winds in the meteor region with one-hour time resolution generally considered optimal. Traditionally, the lidars are meant for a middle atmosphere thermal structure with high time and height resolution. Multiple frequency probing provides wind information with good time and height resolution (She and Yu, 1994). Unlike the MF and Meteor radars, however, lidars seldom provide a long, continuous data record as they are dependent on weather conditions and often require continuous supervision.

As the EPP influence is regional, mainly at auroral latitudes, and with short time scales of up to a few days, ground-based observations are of great importance in studying the dynamical changes. The meteor radars do a good job although the meteor counts are reduced due to ionization (e.g., Pancheva et al., 2007). The meteor radar method for measuring wind assumes that collision frequencies are sufficiently large that the ionized meteor trails assume a bulk motion equal to that of the ambient neutral wind. Kaiser et al. (1969) showed in their theoretical studies that in large electric fields meteor trail can be divided into motions of both the plasma and the ambient neutral atmosphere. High electric fields such as those that occur during geomagnetic disturbances might decouple the meteor trail from the neutral medium (Reid, 1983; Prikryl et al., 1986), leading to erroneous measurements of the neutral wind during sufficiently disturbed conditions. Hocking (2004) reported that there is an anisotropy in the rate of expansion of trails formed above 93 km altitude with a distinct diurnal variation. It has been suggested that this diurnal variation is due to external electric fields that are tidally driven. It is worth noting that both the lidar and meteor wind analyses assume that the vertical wind is zero, which might be violated during strong Joule heating events (Banks, 1977; Price and Jacka, 1991). As the neutral wind impact during these events is of fundamental interest in itself it is, hence, very important to quantify the errors in the winds due to the geomagnetic disturbances. This can be achieved by comparing the meteor radar winds with different remote sensing measurements, such as those of the lidars.

Co-located lidar and meteor wind measurements especially during high ionization periods are rather sparse at auroral latitudes. We were able to inter compare co-located measurements during the declining phase of an SPE (Nesse Tyssøy et al., 2008). Although the meteor radar observations are available during the entire month, the lidar measurements are limited, as only 14 h of measurements during 21–22 January 2005 are available. In this paper, we will investigate the correlation between the two wind measurements in the MLT region. We will assess the correlation between the different zonal wind measurements, as well as discuss potential sources of errors associated with geomagnetic disturbed periods.

This paper is organized as follows: Section 2 describes details of meteor radar and lidar and riometer data, along with a description of the methodology used in this study. Section 3 describes the results from different statistical comparisons between radar and lidar wind measurements and possible reasons for the observed biases and their consequences. Section 4 deals with discussion about the results and section 5 lists the conclusions drawn from the present study.

## 2. Database and analysis

The complementary instrumentation at and near Andøya offers an opportunity to investigate the compatibility of wind measurements based on meteor radar and lidar measurements during different geomagnetic conditions. Both the Skiymet meteor radar and the ALOMAR Weber Na lidar estimate winds in the altitude region 80–100 km. In addition, we will use the cosmic radio noise absorption in selected beams measured by the Imaging Riometer for Ionospheric Studies (IRIS) as a proxy for the electron density variation above Andøya associated with disturbed geomagnetic conditions. A brief description of each technique and its measuring principle is given below.

### 2.1. Observational techniques

#### 2.1.1. Meteor radar

The meteor radar used in this study is located at Andenes (69.27°N, 16.04°E). It is a commercially produced Skiymet radar (Hocking et al., 2001a) designed for all sky real time meteor detection. The meteor radar operates at a frequency of 32.55 MHz with a peak power of 12 kW and transmits radio pulses with a length of 13.3 μs that corresponds to typical sampling resolution of 2 km. At lower elevation angles (less than about 60° (30° from zenith)), the resolution is further degraded due to angular effects - an accuracy in locating the meteor of 1° leads to an additional height error of 1 km or so, so the overall resolution is more than 2 km. The meteor radar system transmits short electromagnetic pulses with a broad polar diagram using one vertically directed three-element Yagi antenna. If the meteor ionization trail is aligned perpendicular to the direction of line of sight from the radar to the meteor, it reflects the transmission signal backwards. The backscattered signal is received by the reception system, which consists of five crossed two element Yagi antennas. The five receiving antennas are arranged in the form of an asymmetric cross, with two perpendicular arms having lengths of 2λ, and the other pair of perpendicular arms having lengths of 2.5λ. Meteor locations are determined from the phase information recorded at the receiving antennas using an interferometric technique with an accuracy of better than ± 1.5-2° (Jones et al., 1998). The meteor detection and discrimination is done through regressive detection algorithms and a detailed description of the detection process can found in Hocking et al. (2001a).

From each specular meteor echo, the radial velocity of the meteor trail due to the projected background wind is estimated. To estimate the horizontal winds, an all-sky least squares fit is applied to the radial velocities of meteors detected within a specific altitude-time window, typically covering a height region of 3–4 km and a time duration of about 1.5 h. The analysis assumes a uniform wind u = (u, v, w) and minimizes the quantity $\sum_{i}(\{u.r_i^u\} - v_{ri})^2$, where $i$ refers to the meteor number in a specified altitude-time window. The vector $r_i^u$ is a unit vector pointing from the radar to the i$^{th}$ meteor trail. The value $v_{ri}$ is the measured radial velocity, and $u.r_i^u$ is a dot-product. In general, the vertical velocities are assumed to be zero. If the difference between measured radial velocity and observed radial velocity is greater than 30 m/s, then the particular meteor is rejected as an outlier. The analysis will be repeated with the meteors that pass the threshold test. The altitude-time window is stepped at time steps of 1 h and height steps of 3 km. In general, the meteors detected at zenith angles between 10° and 60° are used for the horizontal wind estimation in order to avoid overhead reflections and to avoid range ambiguity at higher zenith angles. The horizontal winds are estimated in six height range bins 80.5–83.5, 83.5–86.5, 86.5–89.5, 89.5–92.5, 92.5–95.5, and 95.5–99.5 km and are assigned to 82 km, 85 km, 88 km, 91 km, 94 km and 98 km, respectively.

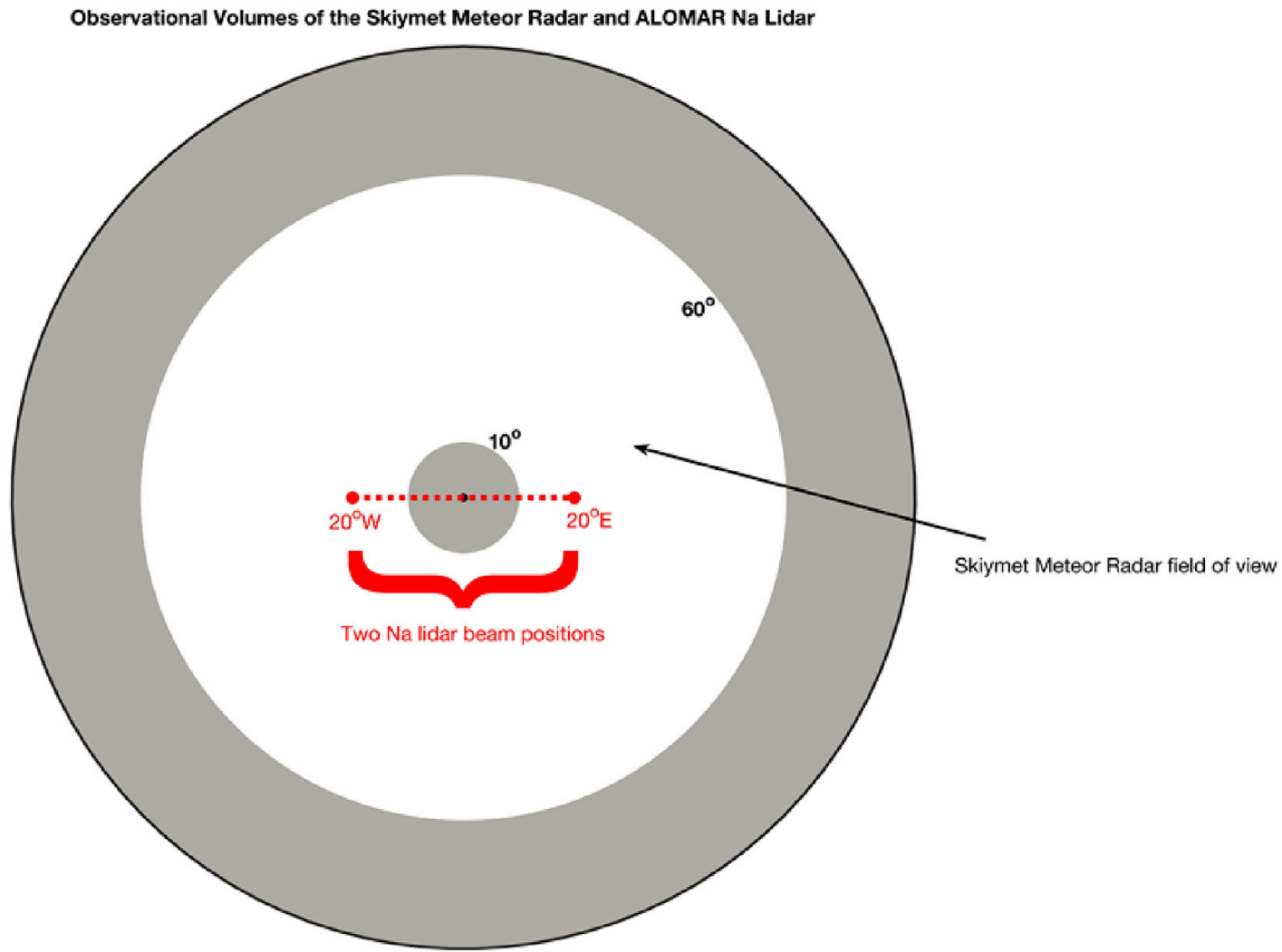

**Fig. 1.** A schematic view of the observational volumes of the Skiymet meteor radar and the Alomar Na lidar. The red dotted line indicates the lidar beam view. The outer boundary indicates the observational area of the meteor radar, while the white area (zenith angles of 10°-60°) indicates the area considered for wind estimates in the present study. (For interpretation of the references to colour in this figure legend, the reader is referred to the Web version of this article.)

### 2.1.2. The ALOMAR Weber Na LIDAR

The ALOMAR wind-temperature Na LIDAR has been part of the Arctic Lidar Observatory for Middle Atmosphere in the period from August 2000 up to June 2017. It is a sodium fluorescence lidar (Vance et al., 1998; She et al., 2002; Arnold and She, 2003). This instrument is used to determine the Na density profiles as well as atmospheric temperature and wind from about 80 to 100 km by remote spectroscopy. The lidar system emits light at three known frequencies, with the center frequency of the D2a line at 589.189 nm and two up and down shifted frequencies at 630 MHz ($\pm$589.189 nm). These three frequencies allow for the theoretical shape of the D2a line to be calculated, providing radial wind and temperature estimates (She and Yu, 1994; She et al., 2002). When the lidar beams are tilted off zenith, horizontal components of the wind field can be obtained by assuming that the vertical wind is zero when averaging over time. A detailed description of the system can be found in Bossert et al. (2014).

During 21–22 January 2005, the Na lidar beams were tilted 20° east and west of zenith, which corresponds to a horizontal separation from overhead of about ± 32 km ( ±0.8° longitude) at 90 km altitude. Hence, assuming the vertical wind to be zero, the zonal winds can be estimated with high vertical (150 m) and temporal (15 min) resolutions for each of the beams. We use the average based on both beams as our final estimate of the zonal wind. During this campaign we chose to use both beams to measure zonal wind more accurately rather than measuring both zonal and meridional wind. The beams are split from the same laser source, so any instrumental wind zero-point offset cancels out when averaging the zonal winds measured by two beams at opposite azimuths. Combining two measurements separated by 64 km also reduces the effect of small-scale wind variations not seen in the radar data.

### 2.1.3. Riometer

An additional data set, which is used to substantiate the ionospheric conditions, is the cosmic radio noise absorption data from the Imaging Riometer for Ionospheric Studies (IRIS) located at Kilpisjarvi, Finland (69.05°N, 20.79°E). Riometers respond to the integrated absorption of cosmic radio-frequency noise of galactic and extra-terrestrial origin through the ionosphere and electron density at heights where there is a high collision. Any deviation from the expected signal reflects the absorption capability and hence the electron density in the ionosphere. Hargreaves (2005) show that during night time most of the absorption takes place at the altitudes 75–85 km. IRIS is operated at 38.2 MHz in 49 beams (Browne et al., 1995). The projection of these beams at 90 km spans the area of 67.8°–70.2° N, and 17.8°–23.8° E. Beam projection of 49 beams can be found in Nesse Tyssøy et al. (2008). While none of the riometer beams overlap directly with the study location, the closest beams, 8 and 15, should be representative of the local absorption above Andøya. We use the level of absorption measured by beam 8 and 15 to determine favourable periods in time where the ionospheric electric field might penetrate deeper into the atmosphere and potentially decouple the meteor ion trail from the background wind.

### 2.2. Methodology

Ideally, two techniques measuring the same quantity will produce sample correlation coefficients of close to unity. The departure from unity will be a function of the experimental errors/limitations of both techniques. Differences in probing methods, spatial and temporal coverage can further cause deviations from an ideal situation. A schematic view of the observational volume of meteor radar and lidar are depicted in Fig. 1. The basic assumption of these two techniques is that the atmospheric variability is homogeneous within the observational volume both in space and time. The lidar wind estimates have high vertical (150 m) and temporal (15 min) resolutions, while the meteor radar wind estimates are obtained at 3 km vertical resolution and 1 h temporal resolution and is an average over almost the entire sky. Hence, the meteor wind estimates show generally less variability compared to the lidar wind estimates, as quasi-random fluctuation due to, for example, gravity waves are being averaged out. The best way to compare these two techniques is to bring both datasets into the same altitude and time resolution. In order to do that, we average the lidar wind estimates to the altitude-time windows of the meteor radar wind estimates. The area of observations is, however, different, giving different observational volumes that might cause significant discrepancies between lidar and meteor radar winds.

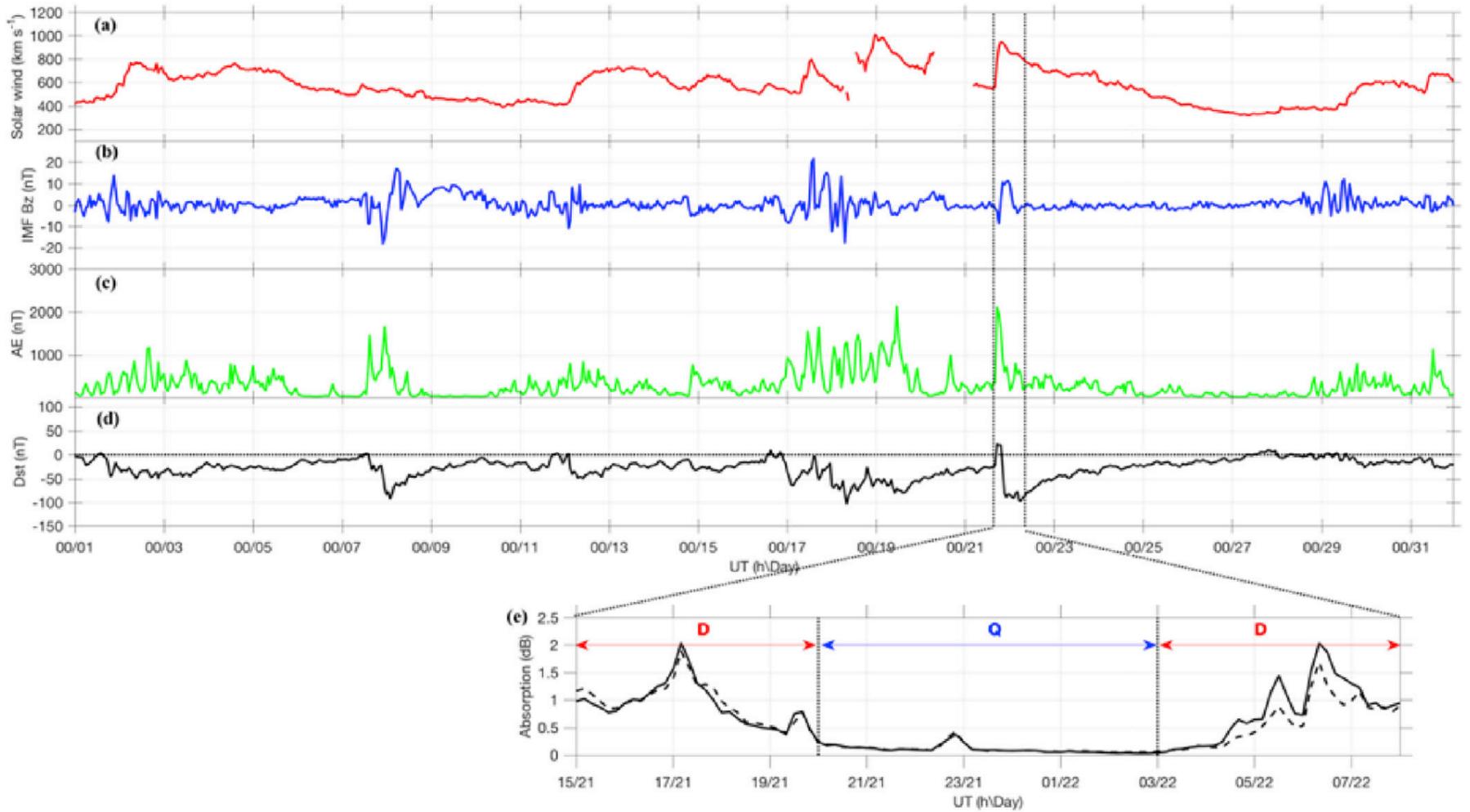

**Fig. 2.** Temporal variation of (a) Solar wind, (b) IMF Bz, (c) Auroral Electrojet (AE) index, (d) Dst index observed during Jan 2005. (e) Cosmic radio noise absorption measured by imaging radio meter at Kilposjårvi from 15 UT, 21 Jan 2005 to 08 UT, 22 Jan 2005. Note the solid line indicates Beam 15 and dotted line indicates Beam 8.

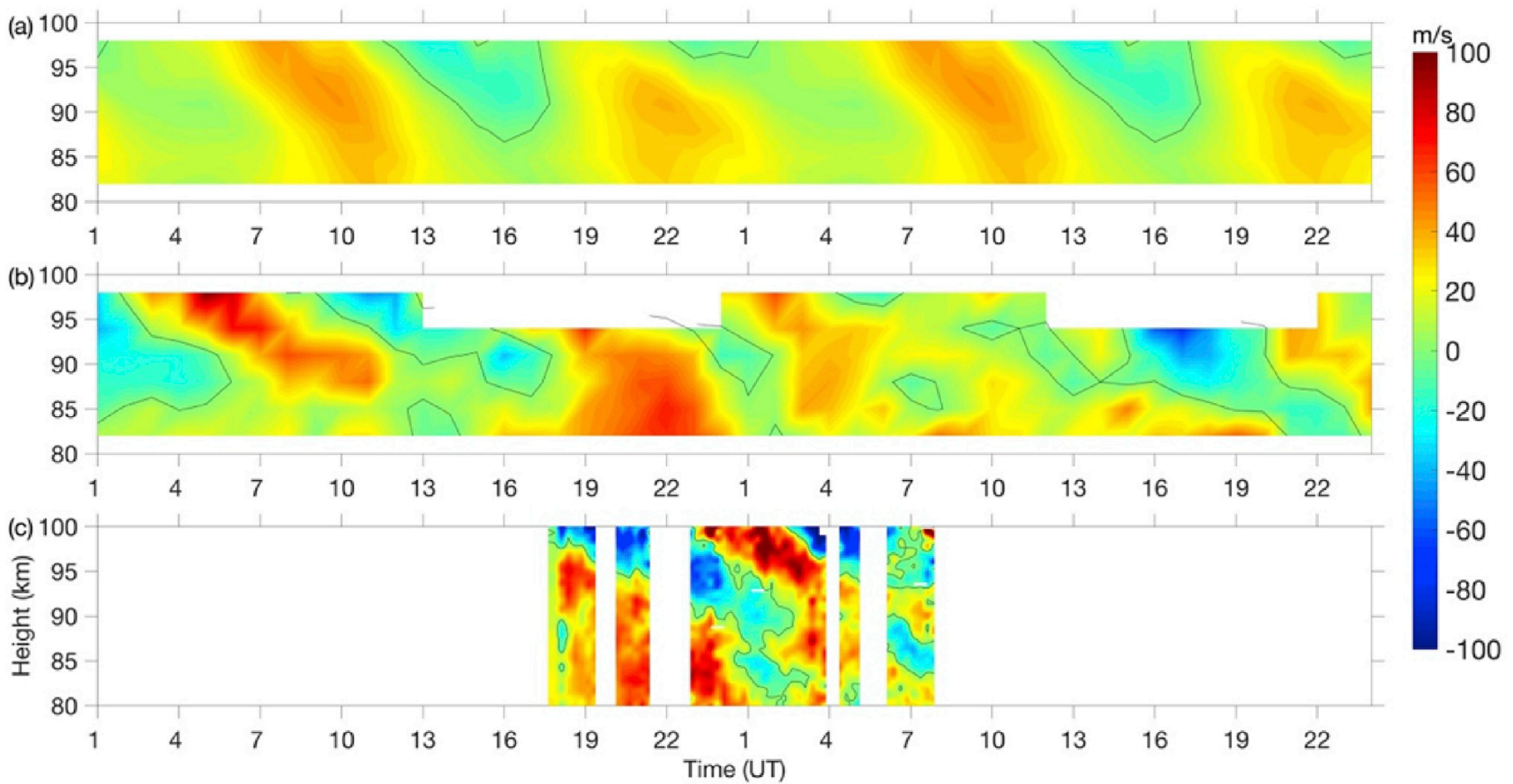

**Fig. 3.** Hourly mean zonal winds (a) composite of entire month of Jan 2005 from meteor radar, (b) 21–22 Jan 2005 from meteor radar and (c) 21-22 Jan 2005 from lidar.

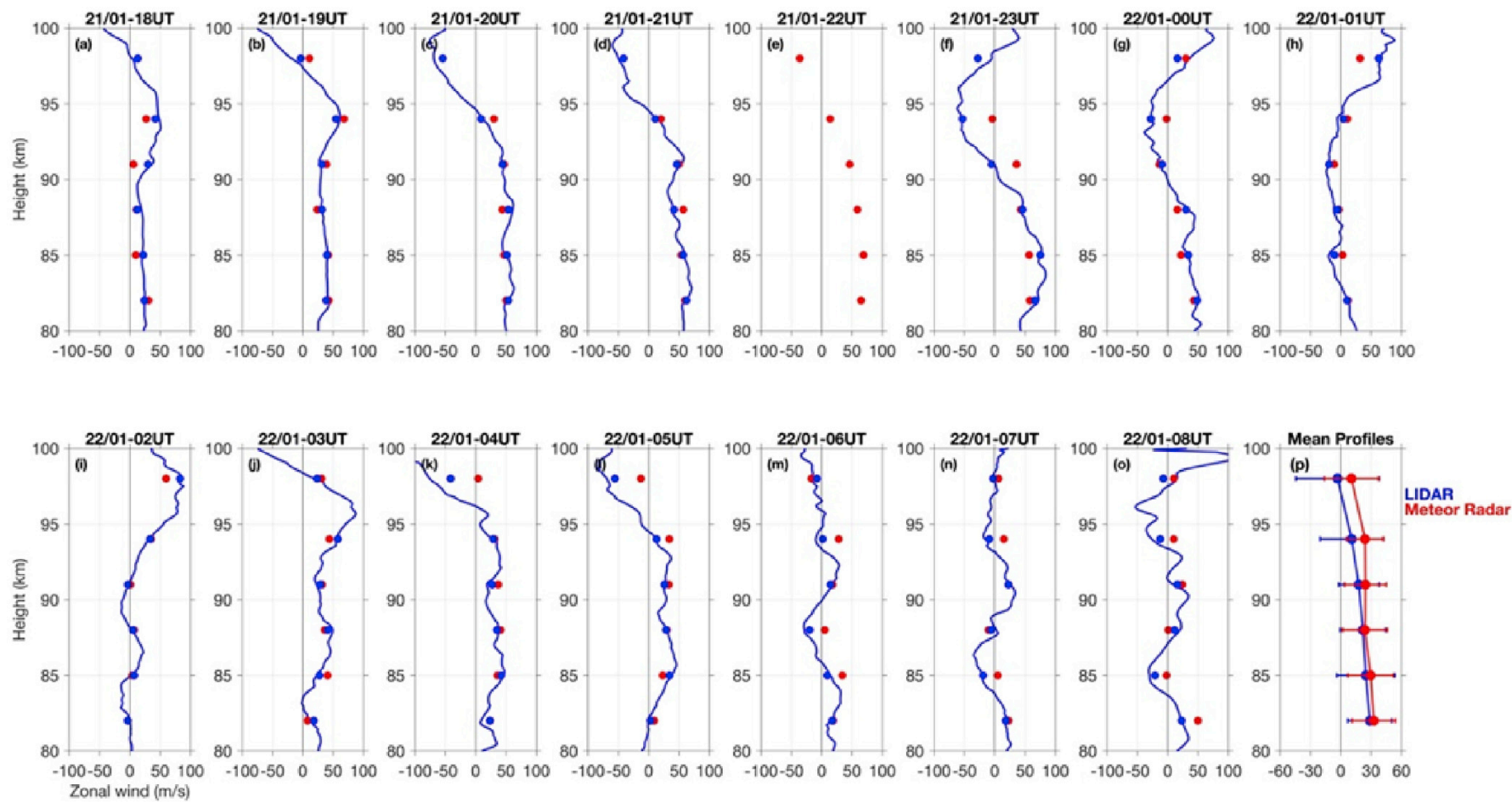

**Fig. 4.** Height profiles of zonal winds from Meteor Radar (red filled circles) and LIDAR (blue line with original height resolution and blue filled circles, averaged over Radar ht. regions) (a-o) for each hour and (p) Mean profiles, estimated from hourly mean winds between 18 UT/21 Jan 2005 and 08 UT/22 Jan 2005, with standard deviation. (For interpretation of the references to colour in this figure legend, the reader is referred to the Web version of this article.)

## 3. Results

### 3.1. Background conditions

Before discussing the details of the results, we briefly describe the background geomagnetic and MLT dynamical conditions during the period covering 21–22 January 2005.

#### 3.1.1. Geomagnetic conditions

Fig. 2(a)-2(d) show the temporal variations of solar wind, The northward component of the Interplanetary Magnetic Field (IMF) Bz, the Auroral Electrojet (AE) index, the Disturbance storm time (Dst) index, respectively, during January 2005. The solar activity in this month was highly variable and clearly evident in the indices. A series of powerful X class flares was produced during this period, resulting in a series of solar proton events (SPEs) starting on 16 January 2005. Complete details of the space weather of this period can be found in the preliminary reports and forecast of Solar Geophysical Data, called "The Weekly" (ftp://ftp.ngdc.noaa.gov/STP/swpc_products/weekly_reports/PRFs_of_SGD/). Importantly, a series of SPEs occurred during this period where the first started on 16 January 2005. Additionally, the aftermath of the SPEs was accompanied by an intense geomagnetic storm and substorm activity. The storm activity is clearly visible in cosmic radio noise absorption measurements, as illustrated in Fig. 2(e).

Fig. 2(e) illustrates 10 min average of cosmic radio noise absorption measured by the riometer operated at radio frequency $f_o = 38.2$ MHz in beam 8 (dotted line) and beam 15 (solid line) from 15 UT, 21 January 2005 to 08 UT, 22 Jan 2005 as first presented in Nesse Tyssøy et al. (2008). The two beams have similar values, which indicates that the particle precipitation is fairly uniform over a large area. We noticed two absorption events during the period. The first one starts around 15 UT on 21 January 2005 and the absorption values decrease to quiet values during 20 UT on the same day. A second absorption event that was observed after 4 UT on 22 January 2005 ended after 12 UT on 22 January 2005. Nesse Tyssøy et al. (2008) reported that the first absorption seems to be dominated by proton fluxes with initial energies similar to precipitating solar protons and the second absorption is associated with enhanced electron fluxes with energies similar to auroral electrons. Based on the absorption data, we divide the data set into Quiet (20 UT on 21 January 2005 to 04 UT on 22 January 2005) and Disturbed (15–20 UT on 21 January 2005 and above 04 UT on 22 January 2005).

#### 3.1.2. MLT region

Fig. 3 shows the zonal winds estimated from both the meteor and lidar measurements. As the meteor radar observations are available around the clock for the entire month, we consider the January 2005 composite hourly mean meteor radar winds as reference. Fig. 3(a) illustrates the composite hourly mean winds for the entire month of January 2005. The meteor radar zonal winds show downward phase propagation with a semidiurnal period, which is the more prominent tidal component over the study location during the winter period (e.g., Pancheva and Mitchell, 2004). The hourly mean meteor radar winds for the period 21–22 January 2005 are illustrated in Fig. 3(b). These winds clearly resemble the semidiurnal variation as in the composite winds but with higher amplitudes. The lidar zonal winds are shown in Fig. 3(c) with its original height resolution (150 m) and with high time resolution (15 min). In all figures, the zero-wind line is highlighted with a black line. Both meteor radar and lidar zonal winds during 21–22 January 2005 resemble the mean composite behavior with slight deviations in amplitudes. The zero-wind line in lidar zonal winds is evident continuously from higher altitudes to lower altitudes, while in the meteor radar winds it is not continuous, at least for the overlap period. As for the wind amplitudes, there are slight differences between lidar and radar winds. More detailed discussion about these discrepancies follows in the next sections.

### 3.2. Radar and lidar wind comparison

Fig. 4(a)-4(o) illustrates the lidar and the radar zonal wind profiles, similar to Fig. 3, but on hourly basis. The lidar zonal winds with the

**Table 1**
Statistical results for Meteor Radar (MR) and lidar zonal winds at all height regions and also individual heights.

| Parameter | All | 82 km | 85 km | 88 km | 91 km | 94 km | 98 km |
| --- | --- | --- | --- | --- | --- | --- | --- |
| Number of coincidences | 80 | 14 | 14 | 14 | 14 | 14 | 10 |
| Correlation coef. | 0.84 ± 0.02 | 0.91 ± 0.03 | 0.89 ± 0.04 | 0.89 ± 0.04 | 0.78 ± 0.07 | 0.88 ± 0.04 | 0.88 ± 0.05 |
| Meteor Radar mean winds | 24 | 30.5 | 26.86 | 21.57 | 22.93 | 24.8 | 15.4 |
| Variance in MR winds | 413 | 413 | 410 | 425 | 417 | 352 | 538 |
| LIDAR mean winds | 19.15 | 28.83 | 24.95 | 21.91 | 17.98 | 10.95 | 6.72 |
| Variance in LIDAR winds | 782 | 470 | 829 | 500 | 403 | 967 | 1794 |
| Mean differences | 3 | 1.1 | −3 | −2 | 3.7 | 17.15 | 11.17 |
| Mean absolute differences | 8.3 | 4.6 | 12.02 | 5.67 | 5.87 | 18.07 | 15.72 |
| Cov($U_{iL}, U_{iR}$) | 479 | 401 | 517 | 408 | 321 | 515 | 860 |
| Cohen's d | 0.21 | 0.08 | 0.08 | 0.02 | 0.24 | 0.56 | 0.26 |
| p (from t-test) | 0.21 | 0.84 | 0.84 | 0.97 | 0.52 | 0.17 | 0.58 |

original vertical resolution (150 m) but averaged over the meteor radar time span are shown as a blue line. Short vertical variations are clearly evident in those profiles and may be due to the presence of the gravity waves. Recently, Bossert et al. (2014) and Nesse Tyssøy et al. (2008) reported mesospheric gravity wave activity using the same lidar measurements. The lidar winds, estimated on the meteor radar altitude-time window, and meteor radar winds are illustrated as blue and red filled circles, respectively. As can be seen from the figure, both the meteor radar and lidar show almost identical zonal wind patterns. A statistical comparison between meteor radar and lidar zonal winds is illustrated in Fig. 4(p). The error bars indicate the standard deviation of hourly mean winds. Overall, the meteor radar mean zonal wind seems to be more eastward compared to the lidar mean zonal winds. The wind differences are about 3 m/s in lower altitudes (below 90 km) and about 8–15 m/s at higher altitudes (above 90 km). The zonal wind differences may come from intrinsic instrument errors as well as natural atmospheric variability, as the two instruments are measuring different volumes of the atmosphere (see Fig. 1). We will discuss the sources of errors in more detail in section 4.

A statistical comparison has been made between the wind fields measured by the radar and the lidar. Table 1 summarizes the statistics including data from all altitudes as well as the individual altitude ranges. Firstly, we applied cross correlation between the two data sets. With the exception of 91 km, high cross correlation coefficient above 0.8 with no time lag are found at all altitudes, except for 91 km. We noticed an unusual time lag of 2 h at 91 km. In order to verify this time lag, we repeated the cross correlation by removing individual points in the time series. From this exercise, we realised that the wind observations at 91 km during 23 UT mainly cause the apparent time lag between the two observations. The large difference between the lidar and the radar wind estimates during 23 UT might be perturbations due to short horizontal wavelength, and might therefore impact the two observational volumes differently. As the meteor radar wind estimates are based on meteor trail echoes randomly distributed over all sky (roughly with diameter of few hundred km), the associated wind estimates will not resolve short wind perturbations less than the radar volume.

From Table 1, the mean radar/lidar zonal wind component differences are 3–5 m/s at lower heights and above 10 m/s at higher heights. This difference is large compared to that observed by Franke et al. (2005). In their comparison study over a tropical station in Maui (20.71°N, 156.26°W), Hawaii, they observed mean differences of less than 1 m/s. Liu et al. (2002) reported wind differences in the order of 1–2 m/s around 86 km and 6 m/s around 93 km over a mid-latitude station Starfire Optical Range (SOR), Kirtland Air Force Base (35°N, 106.5°W). It should, however, be noted that both Franke et al. (2005) and Liu et al. (2002) used significantly longer observational times for comparison. Goldberg et al. (2006) showed (refer to Figs. 16 and 17) good comparison between zonal winds from radar, lidar and rocket during Macwave rocket campaign on 28 Jan 2003, limited to 3–4 h. The comparison between lidar and radar showed good correlation (0.96) and mean difference of 1 m/s.

Table 1 also shows that the variances of lidar zonal wind are larger compared to the variances of the radar zonal wind, which indicates the presence of a high frequency gravity wave component in lidar winds. The Wilcoxon signed-rank sum test (Wilcoxon, 1992) was used to test the hypothesis that observed median differences arise from a population with a nonzero median. The Wilcoxon test indicates that the observed medians are not significantly different from zero at the 5% level for lower altitude regions, while the case is opposite for higher altitudes. Further, to understand the differences between lidar and radar measurements, we estimate the Cohen's distance (Cohen, 1977) between the two measurements. In general, the magnitude of Cohen's distance will explain differences between two populations. Cohen's distance of less than 0.2, 0.5, and 0.8 indicates a small, medium and large difference, respectively. Hence, as shown in Table 1, there is a small difference at lower altitudes while it increases to medium for higher altitudes. Thus, we conclude that the mean and median differences between the two wind estimates are not significantly different from zero for altitudes below 90 km, but above

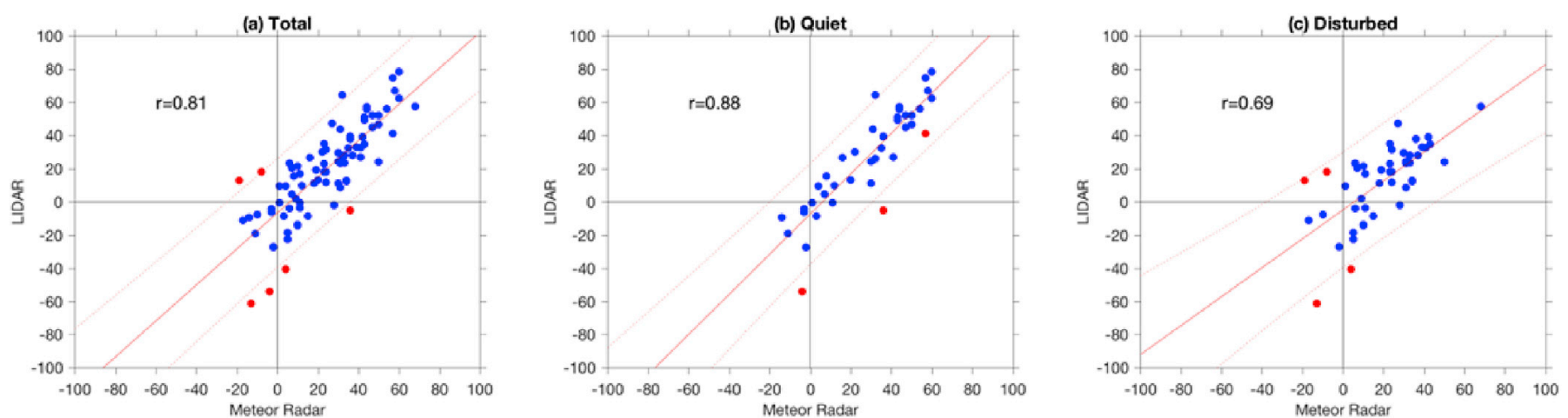

**Fig. 5.** Scatter plots of the zonal wind observed by meteor radar and lidar in the altitude range of 82–98 km (a) Total, (b) Quiet period and (c) Disturbed period. The straight line indicates the linear fit between meteor radar and lidar. The dotted lines indicate the 95% prediction bounds. The red dots indicate the outliers according to cook's distance. (For interpretation of the references to colour in this figure legend, the reader is referred to the Web version of this article.)

Table 2
Statistical analysis results for different cases shown in Fig. 5. Note 'a' stands for slope and 'b' stands for intercept.

|  | Correlation Coefficient | Zero Error in Radar | Zero Error in Lidar | Equal errors in Radar and Lidar ($g_o$) |
|---|---|---|---|---|
| Total | 0.81 ± 0.02 | a = 1.62 ± 0.09<br>b = −5.61 ± 2.63 | a = 0.62 ± 0.05<br>b = 11.49 ± 1.62 | 0.71 |
| Quiet | 0.88 ± 0.02 | a = 0.83 ± 0.13<br>b = −2.98 ± 3.51 | a = 0.57 ± 0.09<br>b = 11.85 ± 2.32 | 0.76 |
| Disturbed | 0.69 ± 0.05 | a = 1.21 ± 0.11<br>b = −7.6 ± 3.8 | a = 0.64 ± 0.06<br>b = 11.12 ± 2.27 | 0.70 |

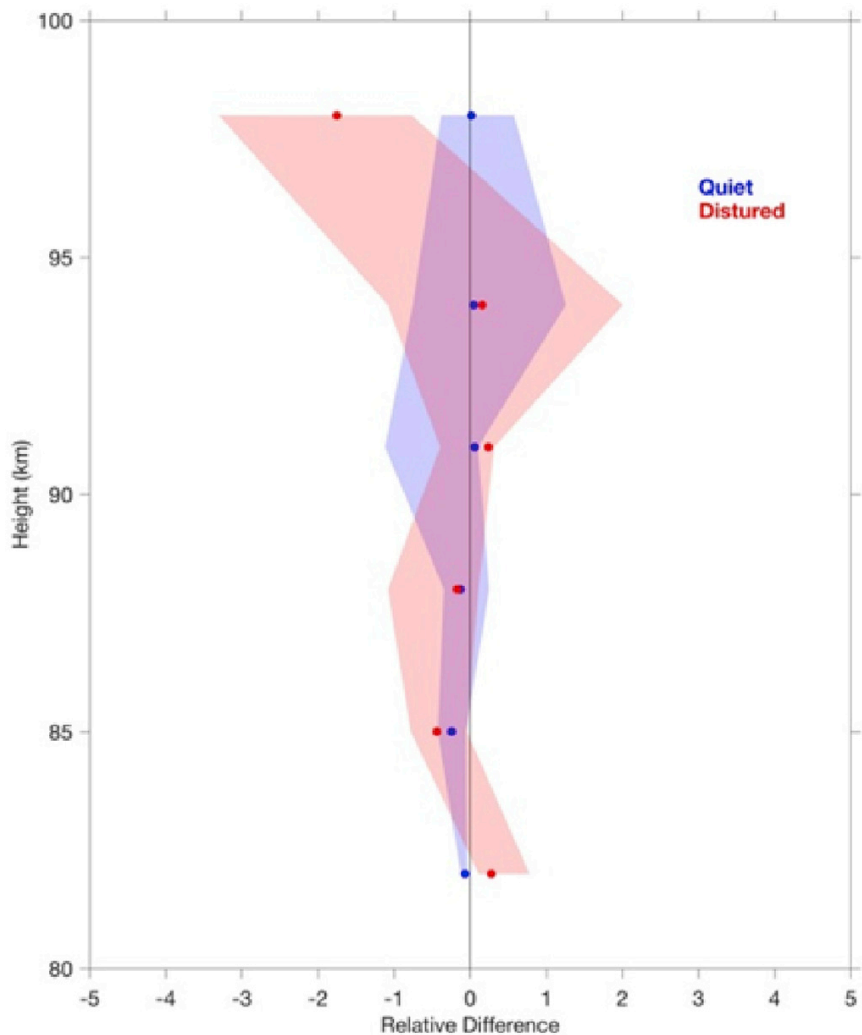

Fig. 6. Vertical profiles of the median relative differences between zonal winds measured by lidar and meteor radar during quite period and distributed period. Shaded region indicates the 25 and 75 percentiles of the relative differences.

90 km there are significant differences between the two wind estimates. At 91 km, however, the statistics are somewhat different from the rest of the altitudes. That might be due to the larger perturbations during one particular hour as mentioned before.

Our analysis so far has revealed a generally strong correlation between lidar and radar zonal wind estimates. Now, we divide the data into two cases, Quiet and Disturbed, based on the absorption data (see Fig. 2(e)) to examine whether there is any influence from the increased electric field and/or ionization impacting the zonal wind estimates. We analyze the meteor radar zonal wind deviations with respect to the lidar zonal winds (see Fig. 5). Fig. 5(a)-5(c) presents the scatter plots for meteor radar zonal winds against lidar zonal winds in the altitude region of 82–98 km including all measurement (total), and for measurement during quiet and disturbed conditions, respectively. We have subjected the data in Fig. 5 to a variety of statistical algorithms. The correlation coefficients with probable errors between the different measurement series are 0.81 ± 0.02, 0.88 ± 0.02 and 0.69 ± 0.05 for Total, Quiet and Disturbed, respectively. The probable errors of correlation coefficients are estimated using an equation $0.6745((1-r^2)/\sqrt{N})$ (Eells, 1929). Here 'r' stands for correlation coefficient and 'N' stands for number of points. The correlation coefficient is improved for quiet conditions compared to total observations, while it is decreased for disturbed conditions. Note that certain points (shown as red dots) biased the data. We estimated these outliers according the Cook's distance (Cook and Weisberg, 1982). Cook's distance is a commonly used estimate of the influential data points when performing a least-squares regression analysis. In general, an observation with Cook's distance larger than three times the mean Cook's distance might be an outlier. We noticed 6 outliers (~5% of total data) in the total time series. All these 6 points occur near the zero-wind line or its adjacent height, which often occur around 91 km and 94 km. As mentioned earlier, the difference in zero wind line would cause large differences between lidar and radar observations. Further, we estimated the regression coefficient between two datasets. In general, the estimation of the regression coefficient assumes error only in one variable and no error in second variable. Hocking et al. (2001b) proposed a least square fit procedure to the datasets assuming both measurements have unknown errors. We applied traditional regression analysis and also method proposed by Hocking et al. (2001b). Details of the regression coefficient and intercept for different cases are listed in Table 2. We also listed the slope of the linear fit ($g_0$) when we assume both measurements have equal errors. From Table 2, the quiet conditions show a larger value of '$g_0$' compared to the other two cases. It indicates that larger biases occur during the disturbed conditions and that is also reflected when considering all measurements not sorted by geomagnetic activity.

Fig. 6 gives the relative differences between lidar and meteor radar winds with respect to the lidar zonal wind as function of altitude. The shaded region in the figure indicates 25 and 75 percentiles. A difference in the relative differences during Quiet and Disturbed conditions is judged significant when the frequency of one sample is not within the confidence interval of the other sample (Wilks, 1995). The difference in the median zonal wind deviation between quiet and disturbed periods is in general small (within the 25 percentile) except for 98 km altitude. The confidence levels using the bootstrap method (Efron and Tibshirani, 1986) show similar results. The large difference at 98 km indicates that there might be considerable errors due to the impact of an increased electric field and/or ionization on the wind estimates.

## 4. Discussion

A limited number of studies have been carried out to understand the dynamical response to the geomagnetic activity. Balsley et al. (1982) reported a correlation between mesospheric zonal neutral winds measured by MST radar at Poker Flat, Alaska and auroral electrojet intensity. Middle atmospheric neutral wind measurements with rocket soundings have also shown considerable changes during geomagnetic activity (Schmidlin et al., 1985). Wand (1983) showed that geomagnetic disturbances cause a 20–25% reduction in semidiurnal tide over Millstone Hill, while Manson and Meek (1991) found a reduction of 10% in the semidiurnal tide due to the geomagnetic disturbances. Recently, Pancheva et al. (2007), Singer et al. (2013) and Trifonov et al. (2016) have shown a dynamical response in MLT winds and tides during the solar proton events using meteor radar observations.

However, to determine the significance of these results and be able to conduct further analysis, we need to understand the reliability of wind meteor radar during these conditions. The basic question "*Are the meteor radar winds reliable during high geomagnetic activity?*" needs to be answered. Previous measurements from Reid (1983); Prikryl et al. (1986) urged that ionization might impact on meteor trails which, in turn, influence wind measurements. Forbes et al. (2001) addressed this problem using meteor radar measurements near 95 km over the South Pole with overhead F region drifts measured by SuperDARN radar at Halley (76°S, 27°W). From their study, they concluded that the electric field contamination of the s = 0 and s = 1 component of the neutral wind field derived from hourly meteor radar data over the South pole is negligible. However, they were not able to discern whether the meteor trail drift is due to a neutral wind or an electric field with their measurements. Hocking (2004) reported that there is an anisotropy in the rate of expansion of trails formed above 93 km altitude with a distinct diurnal variation. It is suggested that this diurnal variation is due to external electric fields that are tidally driven. John et al. (2011) compared the meteor radar winds and TIDI winds over Thumba equatorial station (8.5° N, 77° E) and concluded that Equatorial Electro Jet (EEJ) does not bias the radar measurements.

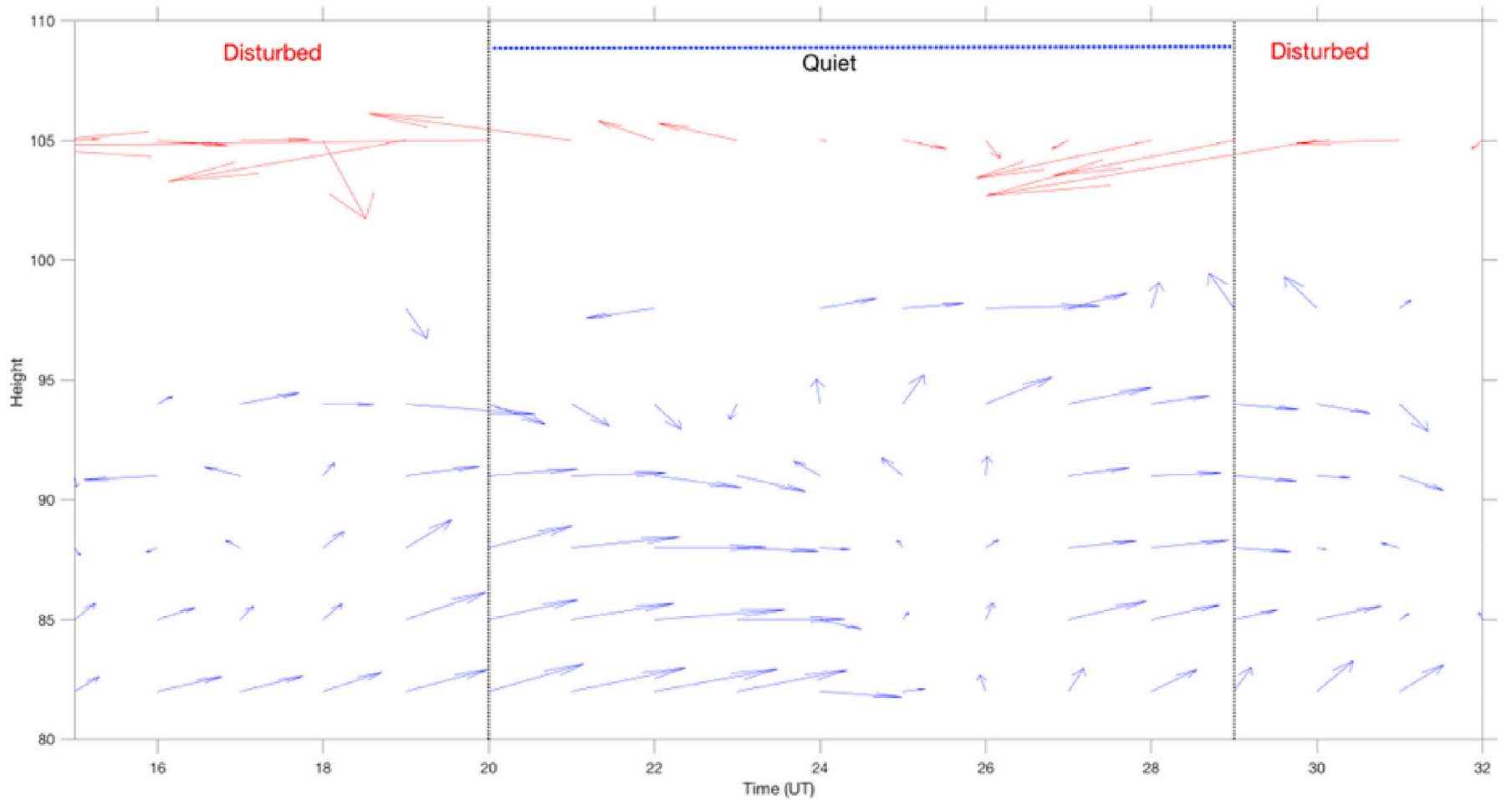

**Fig. 7.** Horizontal wind vectors (blue) from meteor radar observations and auroral drift motions (red) estimated based on magnetometer data from SuperMAG (Gjerloev, 2012) and Andøya magnetometer station. Note that the auroral drift motions are corresponds to the altitude of approximately 105 km and is given in arbitrary units. (For interpretation of the references to colour in this figure legend, the reader is referred to the Web version of this article.)

In the present study, we have made a comparison between lidar and meteor radar zonal winds at Andøya during 21–22 January 2005, coinciding with a period of strong geomagnetic activity in the aftermath of a series of solar proton events. The aim of this study is to investigate whether or not enhanced geomagnetic activity appears to affect the reliability of the meteor radar wind measurements. The comparison between lidar and meteor radar zonal winds shows in general good agreement. The correlation coefficients are around 0.8. Considering the fact that each instrument is measuring different spatial averages, in a wind field that is highly variable both spatially and temporally, we consider these correlation coefficients to be satisfactory. Previous studies by Liu et al. (2002) and Franke et al. (2005) support our findings. We noticed, however, large differences near the zero-wind line. These deviations are expected due to the presence of different wave components in the lidar wind estimates compared to the meteor wind estimates. We found that the long period waves are nicely matched in both winds (figure not shown). This indicates that the gravity wave component in lidar winds due to difference spatial average might cause the large differences near zero wind line. The presence of tidal components cannot, however, be ruled out. The other possible explanation for these discrepancies could be the influence of the geomagnetic disturbances due to particle precipitation. Previous studies by Liu et al. (2002) and Franke et al. (2005) are far from the auroral region and hence free from this effect.

The relative differences between lidar and meteor radar zonal winds during the disturbed conditions are significantly larger compared to quiet conditions, as illustrated in the correlation analysis in Fig. 5 as well as the vertical profiles in Fig. 6. The relative differences fall within each other's distribution at lower altitudes, however, the scenario is totally different at higher altitudes. This implies that changes in winds observed at 98 km might wrongfully attribute apparent wind anomalies during geomagnetic disturbances as true wind changes, whereas it is rather a limitation associated with measurement and/or analyzing techniques.

Although a detailed analysis on the physical mechanism concerning how the strong ionization and/or the associated electric field could influence the measurements or analyzing techniques is beyond the scope of this paper, we consider the following mechanisms as potential candidates. Large electric fields during geomagnetic disturbed conditions may decouple the meteor trail electron motions from the background neutral wind and lead to erroneous neutral winds estimation using the meteor radar Kaiser et al. (1969). Fig. 7 shows the total horizontal meteor wind vectors together with the plasma convection estimated based on magnetometer data from SuperMAG (Gjerloev, 2012) and Andøya magnetometer station. Note that the plasma convection corresponds to the altitude of approximately 105 km and is given in arbitrary units. In general, there are large relative differences between the plasma flow directions and neutral wind directions. During active auroral periods, the meteor radar wind vectors are highly scattered at higher altitudes. Similar results are reported in Prikryl et al. (1986).

Another possibility for large relative differences could be geomagnetic field influence on lidar measurements, called the Hanle effect. Hanle (1924) observed the variation of polarization of the resonance fluorescent light emitted from an atom when it is subject to a magnetic field in a particular direction. A weak magnetic field causes slow Larmor precession, so the dipoles have no time to change their orientation before they spontaneously decay. Consequently, re-emitted fluorescence preserves the polarization of the incident excitation light. On the other hand, in a high magnetic field, fast precession causes rapid averaging of the dipoles orientation, i.e., there is an efficient depolarization of the re-emitted light. Because of the polarization of laser and the magnitude of the Earth's geomagnetic field, the Hanle effect can change the relative intensity of six sodium D2 hyperfine transitions introducing an offset in the Na lidar measurements. Previous observations (Papen et al., 1995; Krueger et al., 2015) using Colorado State University (CSU) lidar reveal that the Hanle effect will have a temperature bias of 1.4 K and a radial wind velocity bias of 0.7 m/s. Although the present lidar system is similar to the CSU lidar, the biases depend on the geographic location and polarization of lidar. *Fricke and von Zahn* [1985] noticed a 5 K systematic temperature shift over Andøya. Note that the biases can vary case to case as they depend on the geomagnetic field strength and orientation.

The third possibility could be excessive Joule heating associated with strong electric fields and enhanced ionization. Joule heating might cause adiabatic upwelling (Price and Jacka, 1991), violating the assumption of zero vertical winds in the data analysis for both the lidar and meteor winds. However, the discrepancies could be a combination of any of these possibilities or might be another unclear mechanism.

## 5. Conclusion

To our knowledge, a comparison between lidar and meteor radar at high latitudes during quiet and disturbed conditions has not been done previously and thus the present study is the first of its kind. Conclusions drawn from the present study are summarized as follows:

➢ Large deviations between lidar and radar winds are found at most higher altitudes mainly during the periods of elevated ionization. These differences might be due to a geomagnetic impact on the observations.
➢ It is hard to attribute the errors to one system. It might be a combination of errors in both systems. We also acknowledge that one event is not sufficient to make any firm conclusion. Further studies with more co-located measurements are needed to test statistical significance of the present findings, as well as investigate the physical mechanism responsible for the discrepancies between the radar and lidar wind estimates.
➢ The chief conclusion that can be drawn from this study is that care must be taken while considering the neutral winds at higher altitudes during geomagnetic disturbed conditions.

### Acknowledgements

This study was supported by the Research Council of Norway under contract 223252/F50. The lidar observations are supported by NSF AGS-1136269. We thank Leibniz-Institute of Atmospheric Physics, Germany for the meteor radar observations over Andøya. For the ground magnetometer data we gratefully acknowledge SuperMAG and its collaborators (http://supermag.jhuapl.edu/). Special thanks to Dr. Gunter Stober to enrich the discussion. We would like to thank the two anonymous referees for help in evaluating this paper.